\documentclass[preprint2]{aastex}

\newcommand{\gap}{\;\rlap{\lower 2.5pt \hbox{$\sim$}}\raise 1.5pt\hbox{$>$}\;}
\newcommand{\lap}{\;\rlap{\lower 2.5pt \hbox{$\sim$}}\raise 1.5pt\hbox{$<$}\;}
\newcommand{\beq}{\begin{equation}}
\newcommand{\eeq}{\end{equation}}

\shorttitle{Binary Black Holes}
\shortauthors{Berczik, Merritt \& Spurzem}

\begin{document}

\title{Long-Term Evolution of Massive Black Hole Binaries. II. \\
Binary Evolution in Low-Density Galaxies}

\author{Peter Berczik \altaffilmark{1,2,3},
 David Merritt \altaffilmark{2}
and Rainer Spurzem \altaffilmark{3}}
\altaffiltext{1}{Main Astronomical Observatory, National Academy of 
Sciences of Ukraine, Kiev, Ukraine}
\altaffiltext{2}{Department of Physics, Rochester Institute of Technology,
Rochester, NY}
\altaffiltext{3}{Astronomisches Rechen-Institut, Zentrum f\"ur Astronomie, Universit\"at Heidelberg, Heidelberg, Germany}

\begin{abstract}

We use direct-summation $N$-body integrations to follow the evolution
of binary black holes at the centers of galaxy models with large,
constant-density cores.
Particle numbers as large as $0.4\times 10^6$ are considered.
The results are compared with the predictions of loss-cone theory,
under the assumption that the supply of stars to the binary is
limited by the rate at which they can be scattered into the
binary's influence sphere by gravitational encounters.
The agreement between theory and simulation is quite good;
in particular, we are able to quantitatively explain the observed
dependence of binary hardening rate on $N$.
We do not verify the recent claim of Chatterjee, Hernquist \& Loeb
(2003) that the hardening rate of the  binary
stabilizes when $N$ exceeds a particular value, or that
Brownian wandering of the binary has a significant effect
on its evolution.
When scaled to real galaxies, our results suggest that massive
black hole binaries in gas-poor nuclei would
be unlikely to reach gravitational-wave coalescence in a Hubble time.

\end{abstract}


\section{Introduction}

Binary supermassive black holes are inevitable
by-products of galaxy mergers, and their 
coalescence is potentially the strongest source of gravitational
waves in the universe \citep{Thorne:76}.
Following their initial formation at separations
of roughly a parsec, massive binaries are expected
to ``harden'' as the two black holes exchange angular
momentum with stars or gas in the host galaxy's nucleus.
The binary separation must decrease by 1-2
orders of magnitude if the black holes are to come close
enough together that gravitational wave emission can
induce coalescence in a Hubble time.
Whether this happens in most galaxies,
or whether uncoalesced binaries are the norm,
is currently an unanswered question \citep{Living:04}.

This paper is the second in a series investigating
the long-term evolution of binary black holes in galactic nuclei.
As in Paper I \citep{MM:03}, we restrict our attention 
to galaxies without gas.
In such an environment, a massive binary shrinks 
as passing stars extract energy and angular momentum 
from it via the gravitational slingshot
\citep{Saslaw:74}.
Early treatments of binary evolution 
\citep{Baranov:84,Mikkola:92,Romani:95,Quinlan:96}
represented the galaxy as fixed in its
properties as the binary evolved, 
and inferred the binary's evolution
from rate coefficients derived via three-body 
scattering experiments.
This approximation is reasonable during the
binary's initial evolution,
but once the separation has decreased by 
a factor of order unity, the assumption of a fixed
background is no longer valid.
The binary quickly (in a galaxy crossing time)
interacts with and ejects most of the stars on intersecting orbits,
and any subsequent binary-star interactions require a
repopulation of the orbits in the binary's ``loss cone.''
This argument has motivated various hybrid approaches to
binary evolution, in which a model for loss-cone
repopulation is coupled with rate coefficients 
derived from scattering experiments in a fixed background
\citep{Zier:01,Yu:02,MM:03,Poon:04,Wang:05}. 

Two regimes of loss-cone repopulation can be identified (Paper I).
If the time scale for encounters (or some other process)
to drive stars into the binary is short compared with orbital
periods, the binary's loss cone will remain nearly ``full,'' 
and the rate of supply of stars will hardly be affected 
by their loss due to ejections.
On the other hand, if time scales for loss-cone repopulation
are long compared with orbital periods,
the binary's loss cone will be nearly ``empty,''
and the binary's evolution will be
limited by the rate at which new stars can diffuse
onto loss-cone orbits.
If the diffusion is driven by star-star gravitational
encounters, 
the binary's hardening rate in this regime
will scale approximately as $N^{-1}$ 
(in a galaxy with fixed mass and radius).
For values of $N$ characteristic of real galaxies,
$N\gap 10^{10}$, encounters would 
be rare enough that the loss cone of a massive
binary would remain nearly empty,
and the hardening rate would correspondingly be very low.
In effect, the decay would stall.

%However we stress that this conclusion is based
%on a highly simplified theory and is 
%guaranteed to be valid only in spherical galaxies 
%with nearly steady-state distribution functions;
%in real galaxies, a number of mechanisms might
%act to repopulate loss cones even if the relaxation
%time is formally very long (Paper I).

While the hybrid models are informative,
a fully self-consistent, $N$-body approach to the
evolution of massive binaries is clearly desirable.
Scattering experiments in a homogeneous background
do not faithfully reproduce the interactions that
take place at the bottom of a galactic potential well,
where a given star may interact more than once
with the binary \citep{Kandrup:03,MM:03}.
If the goal is to follow the 
evolution starting from the pre-merger phase, 
when the two black holes were widely separated,
$N$-body techniques are unavoidable.
However most $N$-body simulations of binary evolution published
to date \citep{Ebisuzaki:91,Makino:93,Colpi:94,MM:01} 
have been based on such small particle numbers that
the binary's loss cone was kept essentially full
by star-star scattering or by random motion of the binary.
As a consequence, these simulations failed to reproduce
the diffusive loss cone repopulation that would
characterize binary evolution in the large-$N$ limit
and they can not easily be scaled to real galaxies.

In this paper, we consider the $N$-body evolution of a massive
binary in a very idealized galaxy model, with a \cite{Plummer:11} 
density profile.
The Plummer model has a large, constant-density core, rather
different from the power-law nuclei of most galaxies.
There are several reasons for this unphysical
choice.
(1) 
The low central concentration of the Plummer model implies
a long star-star relaxation time, hence our $N$-body models
are able to maintain an empty loss cone with fewer particles
than would be required if we had used a more realistic galaxy model.
This allows us to approach more closely than heretofore
to the diffusive loss-cone repopulation regime.
(2) 
Also because of its low central concentration,
the Plummer model evolves only slightly due to the influence
of the binary.
This makes it easier to compare our $N$-body results with
a model in which the gross properties of the galaxy
are assumed to remain fixed with time.
(3) Our initial conditions are precisely the same as those
adopted by Chatterjee, Hernquist \& Loeb (2003) in their
numerical study of binary evolution.
These authors used a hybrid $N$-body code in order to 
achieve large particle numbers
and reached a striking, counter-intuitive conclusion
about the $N$-dependence of the binary hardening rate.
Based on this result, Chatterjee et al. concluded that
``a substantial fraction of all massive binaries in galaxies 
can coalesce within a Hubble time.''
As discussed below, we fail to confirm their result
with our higher-accuracy integrations, and reach
a different conclusion about the likelihood of 
coalescence.

The $N$-body integrations presented here 
were the first to be carried out on a new, special-purpose 
computer that
couples GRAPE hardware with a parallel architecture.
\S 2 describes the computer, the $N$-body algorithm and the 
tests which we carried out to verify its accuracy.
The evolution of the binary is described in \S3,
and in \S4 we show how the $N$-body 
evolution can be reconciled with the
predictions of collisional loss-cone theory.
Our results are compared with those of earlier
studies in \S 5, and \S 6 briefly discusses some
implications of our results for binary evolution
in real galaxies.
\S7 sums up.

\section{Models and Methods}

Our initial galaxy model was a Plummer (1911) sphere, with mass density 
and gravitational potential given by
\begin{mathletters}
\begin{eqnarray}
\rho(r) &=& {3\over 4\pi} {M_{gal}\over r_0^3} \left(1+x^2\right)^{-5/2}, \\
\Psi(r) &=& GM_{gal}\left(1+x^2\right)^{-1/2}, x \equiv r/r_0.
\end{eqnarray}
\end{mathletters}
Here $M_{gal}$ is the total galaxy mass, $r_0$ is the Plummer scale length and
$G$ is the gravitational constant.
We henceforth adopt standard $N$-body units \citep{Heggie:86},
$G=M_{gal}=1$, $E=1/4$ with $E$ the total (binding) energy;
in these units, $r_0=3\pi/16\approx 0.589$.
Particle positions and velocities were generated in the
usual way from the equilibrium, energy-dependent distribution
function $f(E)$.

Our galaxy model was the same one adopted by Chatterjee et al. (2003)
in their $N$-body study of binary evolution.
We also followed their prescription for introducing the massive
binary into the galaxy: two point masses of equal mass, $M_1=M_2=M/2$,
were placed on nearly circular orbits at ${\bf r}$ and $-{\bf r}$, with
$r=0.3$.
No adjustments were made in the stellar positions
and velocities when introducing the black hole particles.
We chose two values for the masses of the black holes:
$M_1=M_2=0.02$ and $M_1=M_2=0.005$.
Integrations were continued until a time of $t=250$,
corresponding to roughly $88$ crossing times, 
where $T_{cr}\equiv (2|E|)^{-3/2}\approx 2.83$.
We carried out integrations with a range of particle numbers,
$N=(0.05,0.1,0.2,0.4)\times 10^6$,
in order to test the dependence of the results on $N$.

The sphere of influence of a (single) black hole of mass $M$
is $r_h\equiv GM/\sigma(0)^2$ with $\sigma(0)$
the 1D stellar velocity dispersion at the center of the galaxy.
In our Plummer spheres, $\sigma(0)=(2/3)\sqrt{2/\pi}\approx 0.532$,
yielding $r_h=(0.141,0.0353)$ for $M=M_1+M_2=(0.04,0.01)$.
The semimajor axis at which a binary first becomes
``hard'' is given approximately by $a_h=GM_2/4\sigma^2$
with $M_2$ the mass of the small component
\citep{Quinlan:96}; for an equal-mass binary, 
$a_h=r_h/8=(1.76\times 10^{-2},4.42\times 10^{-3})$ for
$M=(0.04,0.01)$.
Thus, the two black hole particles moved initially on
widely separated, nearly independent orbits.

The $N$-body integrations were carried out on gravitySimulator,
\footnote[1]{http://www.cs.rit.edu/$\sim$grapecluster/}
a special-purpose computer cluster recently installed 
at the Rochester Institute of Technology.
This cluster contains 32 dual-Xeon nodes running at 3.0 GHz.
Each node hosts a single GRAPE-6A accelerator board \citep{Fukushige:05}
which can store data for up to
131,072 particles and calculate at a speed of 125 GFlops, 
allowing the entire cluster to carry out simulations
with particle numbers of $\sim 4\times 10^6$ 
and at speeds of approximately 4 TFlops.
Communication between the GRAPE boards and their host nodes
is via the standard PCI (32bit/33MHz) interface;
the PC nodes are connected via a high-speed
Infiniband network switch with bandwidth of 10 Gbit/s (duplex).

The $N$-body code was an adaptation of Aarseth's {\tt NBODY1} 
\citep{Aarseth:99} to the GRAPE cluster.
The gravitational force acting on particle $i$ is
\begin{equation}
\mathbf{F}_i = m_i\mathbf{a}_i = -m_i\sum_{k=1,k\ne i}^N 
	{m_k \left(\mathbf{r}_i - \mathbf{r}_k\right) 
	\over \left(\epsilon^2+|\mathbf{r}_i-\mathbf{r}_k|^2\right)^{3/2}}
\label{eq:nbforce}
\end{equation}
where $m_i$ and $\mathbf{r}_i$ are the mass and position
of the $i$th particle and $\epsilon$ is the softening length;
the gravitational force constant has been set to one. 
The integration of particle orbits was based on the fourth-order
Hermite scheme as described by Makino and Aarseth (1992). 
We adopted their formula for computing the time-step of an individual 
particle $i$ at time $t$,
\begin{equation}
\Delta t_{i} = \sqrt{ 
	\eta{|\mathbf{a}(t)| 
	|\mathbf{a}^{(2)}(t)| +
	|\mathbf{\dot a}(t)|^2
	\over |{\bf\dot a}(t)| 
	|\mathbf{a}^{(3)}(t)| + 
	|\mathbf{a}^{(2)}(t)|^2 }}.
\label{eq:ind-ts}
\end{equation}
Here $\mathbf{a}$ is the acceleration of the $i$th particle, 
the superscript $(j)$ denotes
the $j$'th time derivative, and $\eta$ is a dimensionless constant 
that sets the accuracy of the integrator.
Since the particle positions $\mathbf{r}_k$
must be up-to-date in equation (\ref{eq:nbforce}) they are predicted
using a low order polynomial. 
This prediction takes less time if
groups of particles request a new force calculation at large
time intervals, rather than if single particles request
it in small time intervals. 
For this reason, an integer $n$ is chosen such that
\begin{equation}
\left({1\over 2}\right)^{n} \; \le \; \Delta t_{i}
\; < \; \left({1\over 2}\right)^{n-1}
\end{equation}
with $\Delta t_i$ given by equation (\ref{eq:ind-ts}).
The individual time step is replaced by a block time step 
$\Delta_b t_{i}$, where
\begin{equation}
\Delta_b t_{i} = \left({1\over 2}\right)^n.
\end{equation}

The parallel algorithm begins by distributing 
the $N$ particles randomly and evenly between the $p$ 
GRAPEs.
At each time step, the active particles on each node
(i.e. the particles whose positions are due for an update)
are identified and their coordinates broadcast to all of the 
other nodes, where the partial forces are computed.
The partial forces are summed on the head node where the
positions and velocities of the active particles are advanced,
and their coordinates are then updated in the individual
GRAPE memories.
More details of the parallel algorithm, including the
results of extensive performance tests, are given in
an upcoming paper \citep{Harfst:05}.
Most of the integrations used eight nodes; integrations
with the largest particle numbers ($N=0.4\times 10^6$) used either
16 or 32 nodes.

\begin{figure}
\epsscale{0.95}
\plotone{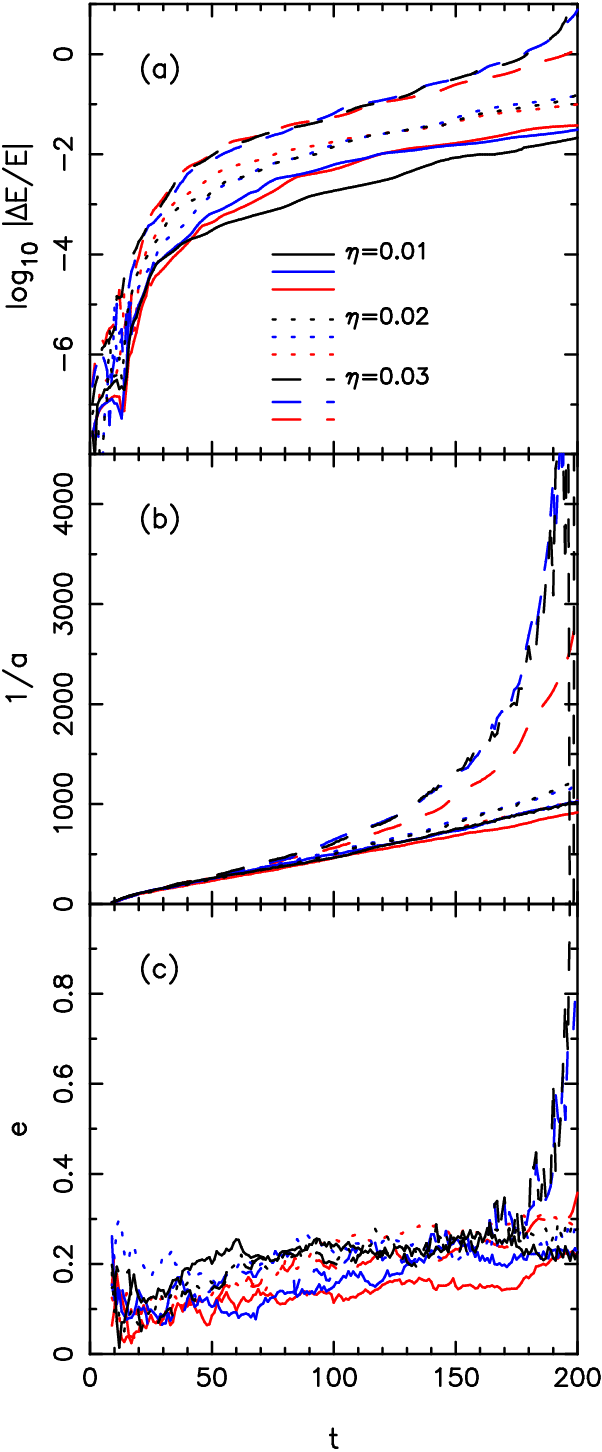}
\caption{\label{fig:perf}
Dependence of the global energy conservation (a),
binary semi-major axis (b), and binary eccentricity (c)
on the $N$-body parameters $\eta$ and $\epsilon$,
for integrations with $M_1=M_2=0.02$ and $N=0.05\times 10^6$.
{\it Black lines:} $\epsilon=1\times 10^{-5}$;
{\it blue lines:} $\epsilon=5\times 10^{-5}$;
{\it red lines:} $\epsilon=1\times 10^{-4}$.}
\end{figure}

The $N$-body code has two important parameters that
affect the accuracy and efficiency of the integrations:
the softening length $\epsilon$ and the time step
parameter $\eta$.
Figure~\ref{fig:perf} shows the effects of varying 
$\epsilon$ and $\eta$
on the evolution of a binary with mass $M_1=M_2=0.02$
in a Plummer-model galaxy with $N=0.05\times 10^6$.
The initial positions and velocities of the two massive particles were the
same as in the ``production'' runs.
The overall accuracy of the calculation as measured by 
changes in energy $|\Delta E/E|$ (Fig.~\ref{fig:perf}(a)) 
depends most strongly on the time-step parameter $\eta$; 
for $\eta=0.01$, the fractional change in $E$ is
a few percent or less.
Figure~\ref{fig:perf}(b) shows that most of the error in the
total energy comes from the binary, which is
very poorly integrated when $\eta$ is as large as $0.03$.
This figure suggests that a value $\eta=0.02$ or
smaller is required for accurate long-term integration of the binary.
The evolution of the binary's eccentricity (Fig.~\ref{fig:perf}(c))
is also poorly reproduced when $\eta=0.03$.
Changing the softening length $\epsilon$ appears to have much less effect 
on the evolution of $a$ or $e$, although of course $\epsilon$
must be small compared with the smallest separation attained by the
binary.
Based on these tests, we adopted $\eta=0.01$ and $\epsilon=1\times 10^{-4}$
for the final integrations.

\section{Evolution of the Binary}

\begin{figure}
\epsscale{1.0}
\plotone{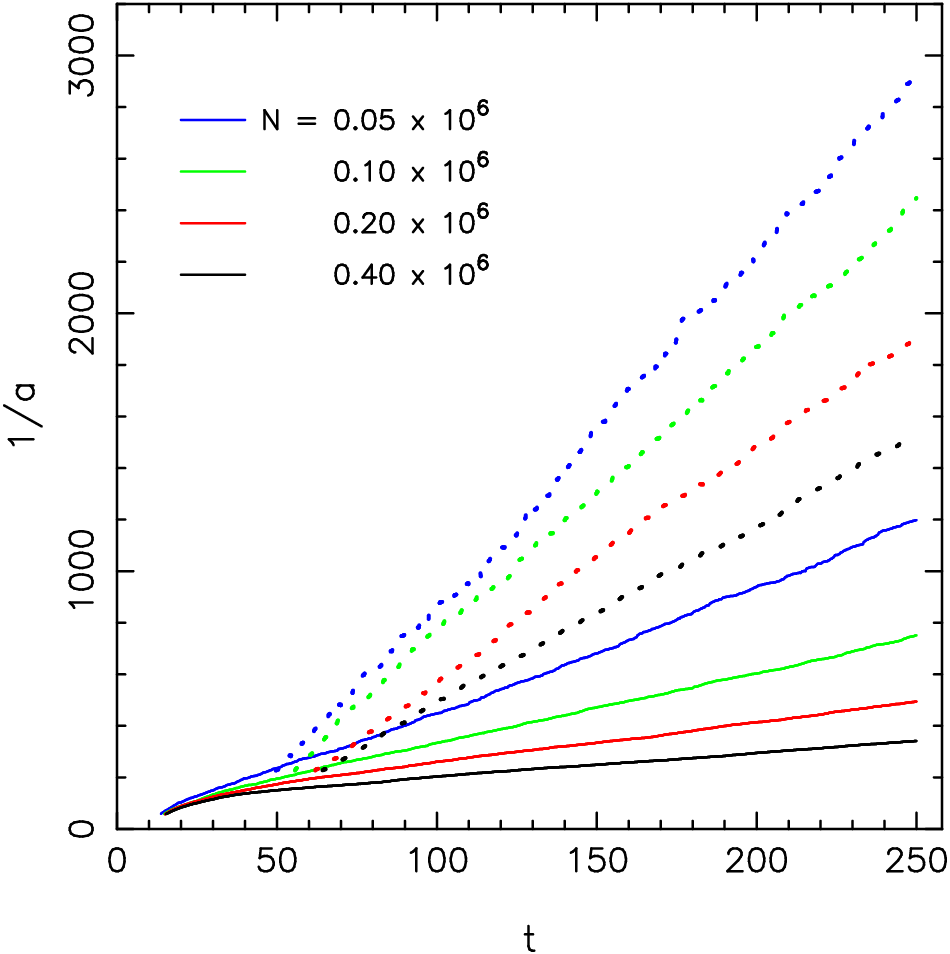}
\caption{\label{fig:ainv}Evolution of binary semi-major axis.
{\it Solid lines}: $M_1=M_2=0.02$; {\it dotted lines}:
$M_1=M_2=0.005$.}
\end{figure}

Figure~\ref{fig:ainv} shows the evolution of the binary's semi-major axis
in each of the eight integrations.
The $N$-dependence of the hardening rate is clear.
At late times, when $a\ll a_h$, the binary semi-major axis obeys
$a(t)^{-1}\approx C_1(N) + C_2(N)t$,
i.e. the binary's binding energy
increases almost linearly with time, and the hardening rate is
a monotonically decreasing function of $N$.
An approximately linear dependence of $1/a$ on time is
characteristic of both the ``full loss cone'' (small $N$) 
and ``empty loss cone'' (large $N$) regimes (Paper I).
If the loss cone were completely full, however, the
hardening rate would be independent of $N$,
clearly a poor description of Figure~\ref{fig:ainv}.

\begin{figure}
\epsscale{1.0}
\plotone{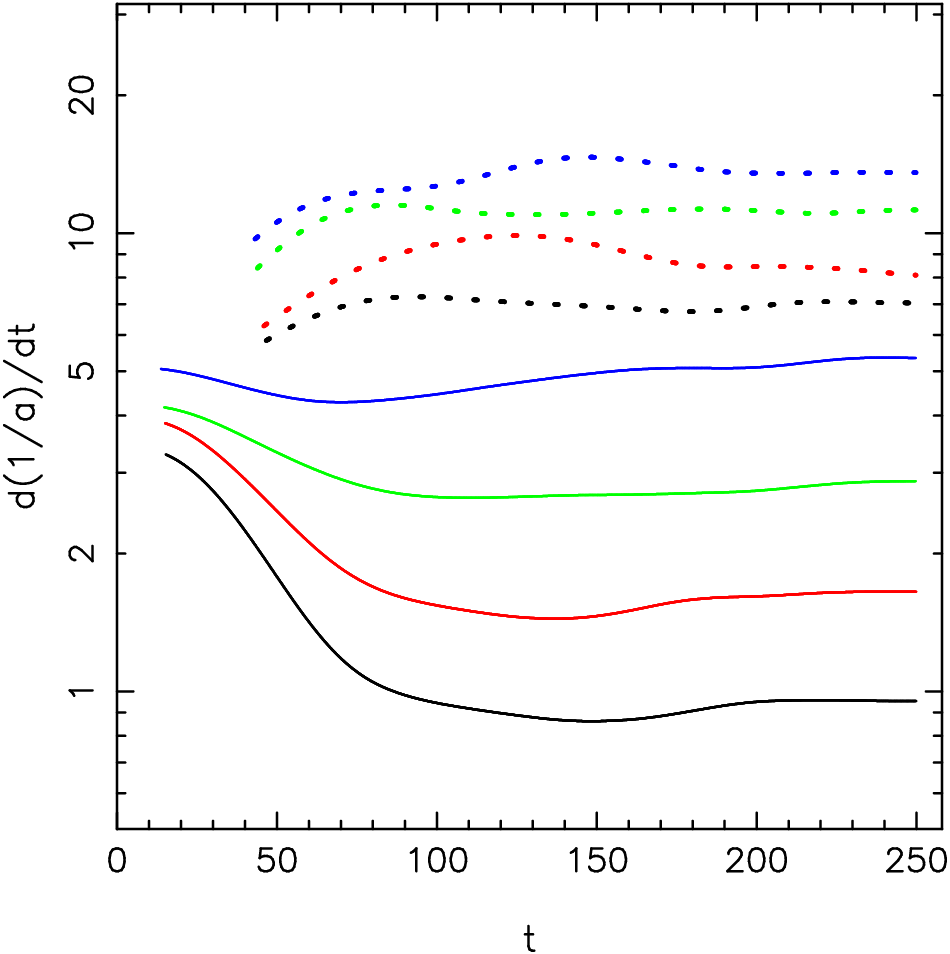}
\caption{\label{fig:slope}Evolution of binary hardening rate.
Color scheme is the same as in Figure~\ref{fig:ainv}.
{\it Solid lines}: $M_1=M_2=0.02$; {\it dotted lines}:
$M_1=M_2=0.005$.}
\end{figure}

\begin{figure}
\epsscale{0.90}
\plotone{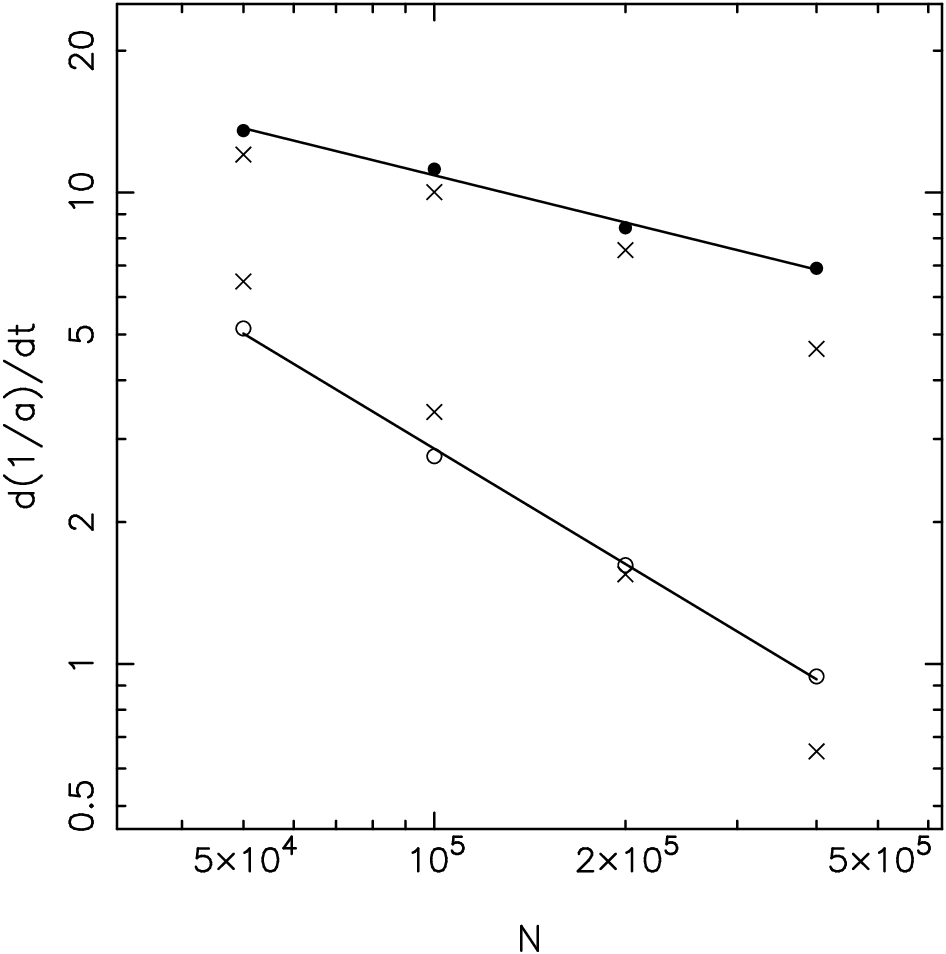}
\caption{\label{fig:ndepend}Mean binary hardening rates in the interval
$150\le t\le 250$ as a function of $N$.
{\it Open circles}: $M_1=M_2=0.02$;
{\it filled circles}: $M_1=M_2=0.005$.
The lines are least-squares fit to the $N$-body hardening rates.
Crosses show the predictions from loss-cone theory, as discussed in
the text.
}
\end{figure}

We define the instantaneous hardening rate to be
\beq
s(t)\equiv {d\over dt}\left({1\over a}\right).
\eeq
Figure~\ref{fig:slope} shows $s(t)$ for each of the integrations,
computed by fitting smoothing splines to $a^{-1}(t)$
and differentiating.
The constancy of $s$ at late times ($t\gap 150$) is apparent;
aside from some wiggles,
the dependence of $1/a$ on time is well approximated as
a linear function.
To a good approximation, we can therefore identify a {\it unique},
late-time hardening rate $\overline{s}(N,M)$ with each of the integrations.
We computed $\overline{s}$ by fitting a
straight line to $a^{-1}(t)$ in the time intervals
$150\le t\le 250$.
Figure~\ref{fig:ndepend} shows the results,
plotted versus particle number $N$.
The $N$-dependence of the mean hardening rate is 
approximately
\begin{mathletters}
\begin{eqnarray}
\log_{10}\overline{s}& \approx & 4.5 - 0.81 \log_{10}N, \ M_1=M_2=0.02, \nonumber \\
                        & \approx & 2.7 - 0.33 \log_{10}N, \ M_1=M_2=0.00, \nonumber
\end{eqnarray}
\end{mathletters}
or
\begin{mathletters}
\begin{eqnarray}
{d\over dt}\left({1\over a}\right) 
&\approx& 3.3\times 10^4N^{-0.81}, \ M_1=M_2=0.02, \nonumber \\
&\approx& 5.0\times 10^2N^{-0.33}, \ M_1=M_2=0.005. \nonumber
\end{eqnarray}
\end{mathletters}

The very nearly linear increase of the binary's binding energy
with time seen in all these simulations was something
of a surprise. 
While $a^{-1}(t)$ is predicted to be
{\it approximately} linear in both the full- and empty-loss-cone
regimes, at least at times before the binary
has removed much mass from the core (Paper I),
the theory on which this prediction is based is only
approximate.
It will be interesting to see if the constant rate of
hardening observed here is a very general feature
of binary evolution.

\begin{figure}
\epsscale{1.0}
\plotone{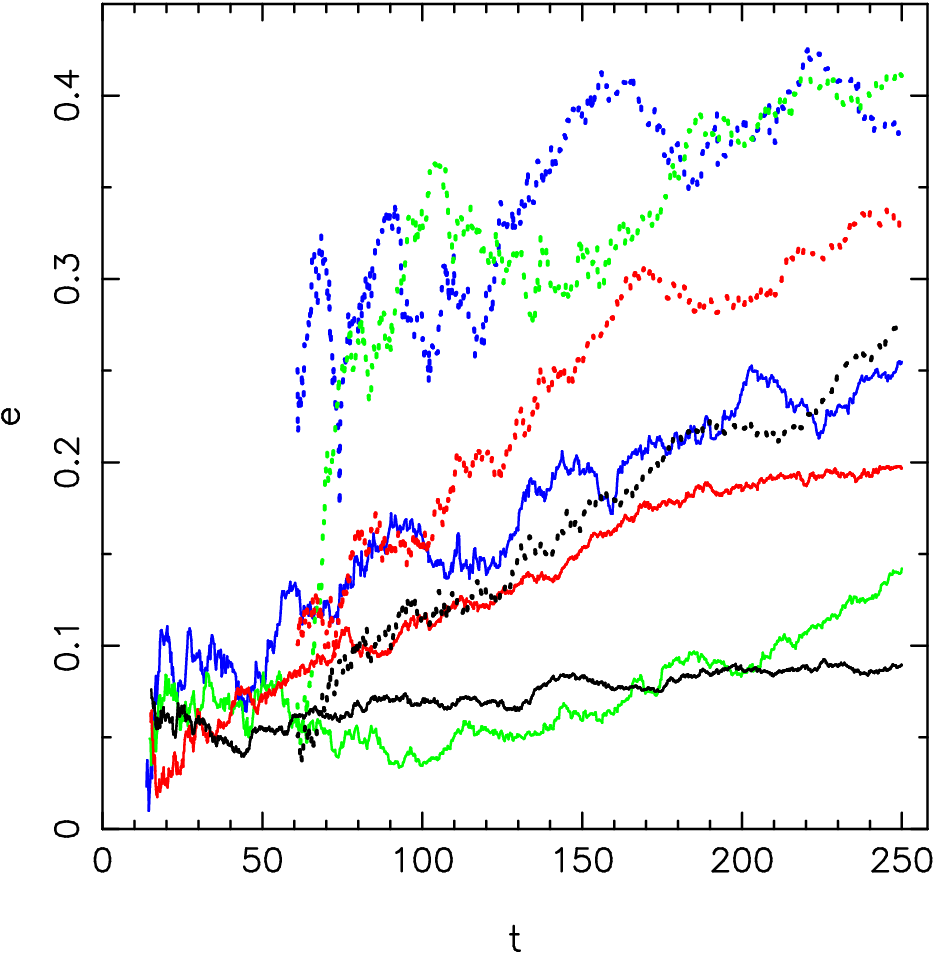}
\caption{\label{fig:ecc}Evolution of binary eccentricity.
Color scheme is the same as in Figure~\ref{fig:ainv}.
{\it Solid lines}: $M_1=M_2=0.02$; {\it dotted lines}:
$M_1=M_2=0.005$.}
\end{figure}

Also of interest is the evolution of the binary's eccentricity
(Figure~\ref{fig:ecc}).
The two massive particles are introduced into the galaxy model
on approximately circular orbits with respect to the galaxy center, but
encounters with stars induce a non-zero eccentricity even before the time
that the separation has fallen to $a_h$, and the eccentricity continues
to evolve as the binary ejects stars.
There is an $N$-dependence here as well, in the sense that
$e$ tends to evolve less with increasing $N$,
although the trend is obscured by noise.
In addition, for larger $N$, the ``initial'' eccentricity
(i.e. the value of $e$ when the binary first becomes hard)
tends to be smaller, due presumably to the smaller size of
random perturbations from passing stars.
Figure~\ref{fig:ecc} suggests that the eccentricity evolution of a binary
in a real galaxy with much larger $N$ would be very small,
although larger particle numbers will be required to
verify this conclusion.

\begin{figure}
\epsscale{1.05}
\plotone{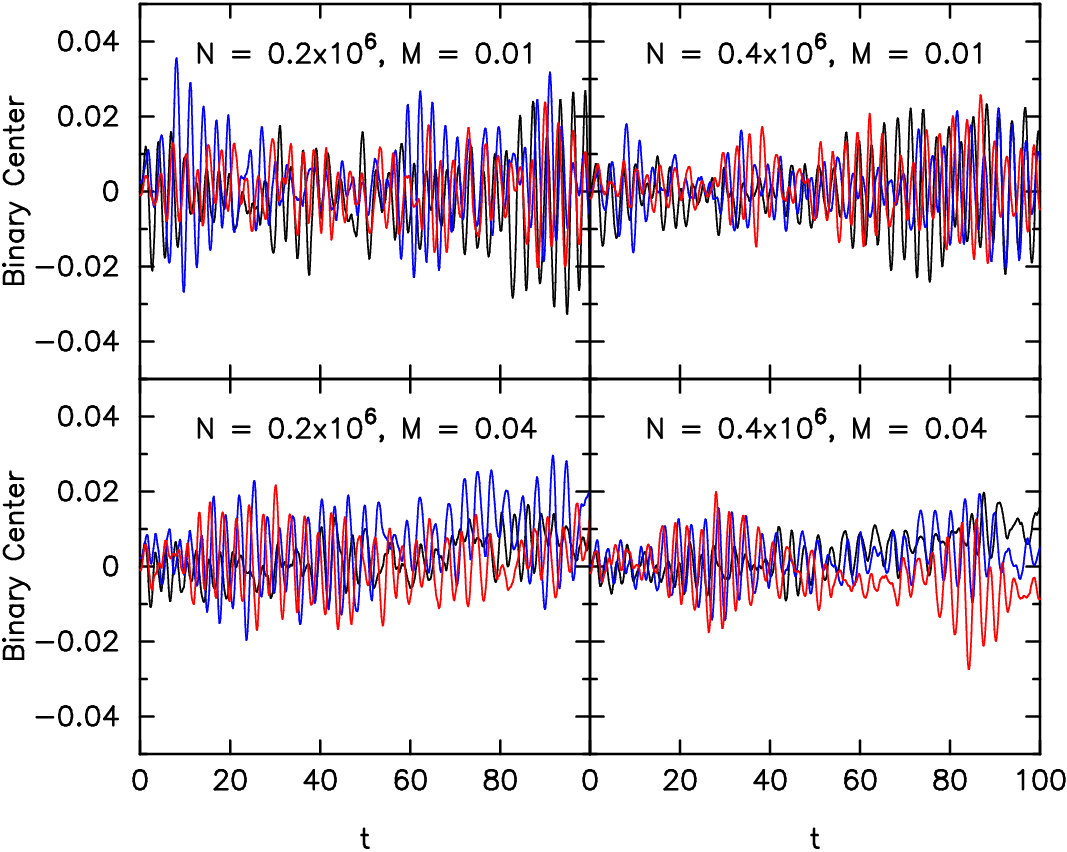}
\caption{\label{fig:brown} Motion of the binary center of
mass in the four integrations with largest $N$.
The position of the binary is measured with respect
to the fixed (inertial) frame; the systematic change 
that is apparent in some of the plots is due to an overall
drift of the $N$-body model.
{\it Black lines}: $x_{CM}$; {\it blue lines}: $y_{CM}$; {\it red lines}: $z_{CM}$.}
\end{figure}

The center of mass of the binary wanders quasi-randomly due
to encounters with stars, both distant, elastic encounters
\citep{CHL:02,Laun:04}
and ``super-elastic'' encounters in which the binary's binding
energy is transformed into linear momentum during ejections
\citep{Merritt:01}.
Figure~\ref{fig:brown} shows this gravitational Brownian
motion in the four integrations with largest $N$.
These plots show the motion of the binary with respect to a
fixed (inertial) frame; because the $N$-body model as a whole
drifts in space, there is also a systematic drift
of the binary's mean position.
We attempted to ``take out''  this drift by computing the
position of the
binary with respect to the galaxy's density center at each time step.
However the structure of the galaxy models, with their
large, constant-density cores, made this difficult
since the position of the estimated density
center tended to vary from time step to time step with an 
amplitude almost as great as that of the
Brownian motion.
In any case, one can estimate the amplitude of the 
``random'' component of the motion from Figure~\ref{fig:brown},
and it is quite similar to what has been found in
other $N$-body simulations with similar initial conditions
\citep{CHL:03,Funato:04}, as well as with the expectations
from encounter theory \citep{Merritt:04}.
Below we address the question of whether the binary's
Brownian motion might influence its hardening rate.

\begin{figure}
\epsscale{1.0}
\plotone{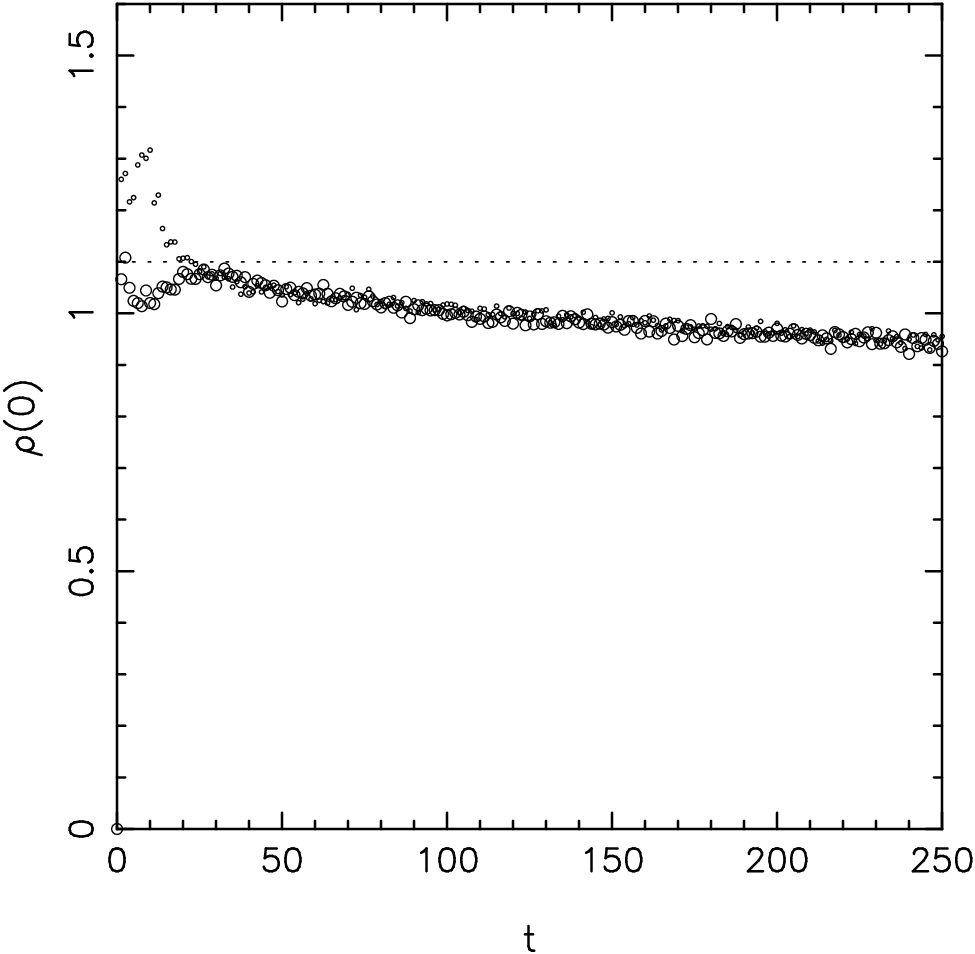}
\caption{\label{fig:rho}Evolution of the central density
of the galaxy,
defined as the mean density within a sphere of radius
$0.2$ centered on the binary,
in the two integrations with $N=0.4\times 10^6$.
{\it Small dots}: $M_1=M_2=0.02$; {\it open circles}:
$M_1=M_2=0.005$.
Dashed line shows the mean density within $r=0.2$ in the
initial model.}
\end{figure}

The presence of the binary affects the central properties
of the galaxy, although only slightly.
Figure~\ref{fig:rho} shows the evolution  of the central density
in the two integrations with $N=0.4\times 10^6$;
here the density was defined as the mass in a sphere of radius 
$0.2$ centered on the binary divided by the volume of the sphere.
The introduction of the binary into the model at $t=0$ causes
an impulsive change in the local density, particularly in the
case of the more massive binary, but the effect is transient
and thereafter the density changes only by $\sim 15\%$ over
the course of the integration, due presumably to ejection
of stars.
The evolution in central density was almost the same for these
two integrations and depended only weakly on $N$ as well.
Such modest changes in the central properties of the galaxy
models could hardly be responsible for the strong dependence 
of binary hardening rate on $M$ and $N$ seen
in Figures~\ref{fig:ainv}-\ref{fig:ndepend}.

\section{A Model for Binary Evolution}

In an infinite, homogeneous background with fixed properties 
(density, velocity dispersion etc.),
the binary's hardening rate is given by
\beq
s\equiv {d\over dt}\left({1\over a}\right) = H {G\rho\over\sigma}
\label{eq:simple}
\eeq
with $\rho$ and $\sigma$ the stellar density and 1D velocity 
dispersion; $H$ is a dimensionless
decay rate that depends on the binary semi-major axis, as well as on
other binary parameters such as mass ratio and eccentricity.
Scattering experiments give $H\approx 16$ for a hard ($a\ll a_h$), 
equal-mass binary \citep{Hills:83,Mikkola:92,Quinlan:96,Merritt:01}.
If we set $\rho$ and $\sigma$ to their values in our Plummer-model
galaxy at $t=0$, equation~\ref{eq:simple} with $H=16$ predicts $s\approx 35$, 
independent of $M$ and $N$.
By comparison, the {\it largest}, late-time decay rate found in these
integrations (for $M=0.01$ and $N=0.05\times 10^6$) 
was $s\approx 14$ (Figures~\ref{fig:slope}, \ref{fig:ndepend}),
not terribly different.
However, the strong dependence of the decay rate on $M$ and $N$ 
in the simulations
is clearly inconsistent with this simple model.

As discussed in Paper I, there is a natural explanation
for the $N$-dependence of the hardening rate.
After roughly a single galaxy crossing time, the binary has ejected
all of the stars on intersecting orbits.
Subsequent interactions between the binary and the stars occur
at a rate that is limited by how quickly new stars can be 
scattered into the binary's loss cone.
The latter is defined as the set of orbits with pericenters
$r_p < K a(t)$, with $K$ a number of order unity; this 
expression reflects the expectation that stars will be ejected
from the binary whenever they pass a distance $\sim a$ from
its center of mass.
But the scattering of stars onto low-angular-momentum
orbits depends on the two-body relaxation time, 
hence on the mean stellar mass, hence on $N$, and so
one expects the hardening rate of the binary to be $N$-dependent as well.

As a first step toward testing this model,
we asked whether star-star encounters might be so frequent
in the $N$-body models
that the binary's loss cone is maintained in a continuously
populated state.
The standard measure of the loss-cone refilling rate
in a spherical, steady-state galaxy is
$q(E)$, defined as
\beq
q(E)\equiv {(\delta J)^2\over J_{lc}^2}.
\label{eq:q}
\eeq
\citep{LS:77}.
Here $\delta J$ is the change over one radial period
in the angular momentum of a star on a low-$J$
orbit, and $J_{lc}$ is the angular
momentum of an orbit at the edge of the loss cone.
A value $q(E)\gg 1$ implies that the loss cone orbits at energy
$E$ are re-populated at a much higher rate than they are de-populated
by the central sink (in this case, the binary black hole), 
and the loss cone remains nearly full.
A value $q\ll 1$ implies that the loss cone is essentially empty,
and repopulation must take place diffusively, as stars
scatter in from $J\gap J_{lc}$.

\begin{figure}
\epsscale{1.1}
\plotone{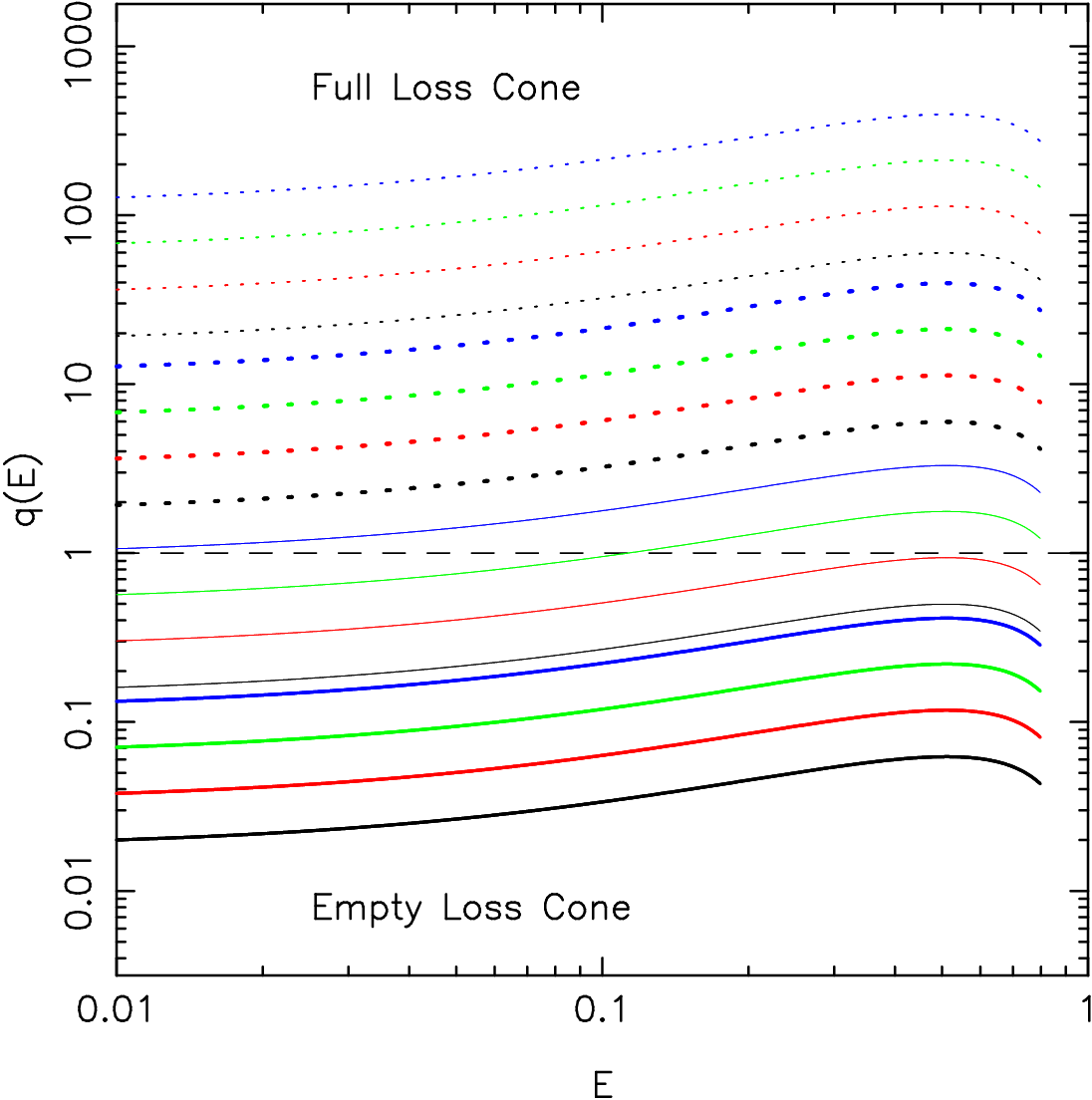}
\caption{\label{fig:qofe}The quantity $q(E)$ that measures the degree
to which the binary's loss cone is repopulated by encounters.
Color scheme is the same as in Fig.~\ref{fig:ainv}.
{\it Solid lines}: $M_1=M_2=0.02$; {\it dotted lines}:
$M_1=M_2=0.005$.
The thicker curves in each group are for $r_t=a_h$, the 
approximate radius of the loss sphere when the binary first
forms, and the thinner curves are for $r_t=a_h/10$, corresponding
to an evolved binary.
These curves suggest that the evolution of the more massive
binary in our $N$-body integrations took place mostly in the 
``empty loss cone'' regime, while the evolution of the less massive binary
took place mostly in the ``full loss cone'' regime.}
\end{figure}

We computed $q(E)$ in our Plummer-model galaxies,
ignoring any effect of the binary on the galaxy.
We used the precise, orbit-averaged definition of $q(E)$ given
by equation (8) in Paper I.
The results are shown in Figure~\ref{fig:qofe}.
The binary parameters appear implicitly in $q$, 
since $J_{lc}^2 = 2GMr_t$ with $r_t=Ka$ the radius of the capture sphere.
Figure~\ref{fig:qofe} shows $q$ as a function of $E$ for the eight different 
combinations of $(N,M)$ in our $N$-body integrations.
We considered two values for $J_{lc}$,
$J_{lc}^2=2GMa_h$ and $J_{lc}^2=2GMa_h/10$;
the first value is the approximate ``size'' of the loss cone
when the hard binary first forms, and the latter value is 
appropriate when $a$ has decreased by a factor of $10$, roughly the 
maximum amount of decay seen in these integrations (Fig.~\ref{fig:ainv}).
Figure~\ref{fig:qofe} suggests that the integrations with the 
higher-mass binary, $M=0.04$, took place mostly in the 
``empty loss cone''regime, $q\lap 1$, 
while the integrations with $M=0.01$ were mostly in the
``full loss cone'' regime, $q\gap 1$.

This result  provides a natural explanation for the different
$N$-dependence of the decay rates in the two sets of integrations.
When the loss cone is empty, $q\ll 1$, re-population takes place
diffusively, as stars are scattered by other stars onto low-angular-momentum
orbits.
The scattering rate scales inversely with the relaxtion time,
or as $\sim 1/N$.
Indeed, the binary decay rate for $M=0.04$ scales almost
as $N^{-1}$ (Fig.~\ref{fig:ndepend}).
On the other hand, when the loss cone is full, $q\gg 1$,
the decay rate is nearly independent of the relaxation time,
since (by assumption)  scattering occurs so quickly that
orbits are never depleted. 
The much weaker dependence of $s$ on $N$ for $M=0.01$
(Fig.~\ref{fig:ndepend}) is qualitatively consistent with this
prediction.

We tested this model more carefully, by computing the expected 
hardening rate of the binary, under the assumption that the supply of stars
was limited by the rate of loss cone repopulation.
The flux of stars into the binary's loss cone (mass per unit time)
in a steady-state spherical galaxy is given approximately by
\beq
{\cal F}(E)dE \approx 4\pi^2J_{lc}^2(E) q(E) {f(E)\over\ln R_0^{-1}}dE
\eeq
(Paper I), where $f(E)$ is the phase-space mass density of
stars at energy $E$, and $R_0(E)$ is the effective, dimensionless
size of the loss
cone as seen by a star of energy $E$; an approximate expression for
$R_0(E)$ is given by \cite{CK:78}.
If all of the stars scattered into the loss cone are assumed
to interact instantaneously with the binary, its decay rate is roughly
\beq
s \equiv \frac{d}{dt}\left(\frac{1}{a}\right) 
\approx {2\langle C\rangle\over aM} \int {\cal F}(E) dE
\label{eq:srate}
\eeq
where $\langle C\rangle$ is the average value of the dimensionless
energy change during a single star-binary encounter,
$C\equiv (M/2m_\star)(\Delta E/E)$ \citep{Hills:83}.
In the limit $a\ll a_h$, $\langle C\rangle$ 
is constant and roughly equal to one.
While equation (\ref{eq:srate}) with constant $\langle C\rangle$
is unlikely to give a good description of the early evolution of the 
binary, when $a\lap a_h$, it should reproduce the late-time 
hardening rate of the binary fairly well.
Figure~\ref{fig:ndepend} shows that this is indeed the case;
we set $r_t=a$ and $\langle C\rangle=1.25$, 
and evaluated $s$ from equation (\ref{eq:srate})
at the final value of $a$ reached in each integration.
Given the simplicity of the model, the agreement with the
$N$-body hardening rates is remarkably good.
Deviations between theory and simulation appear to be greatest at largest $N$; 
this may be due to the fact that the binary separation
does not fall far below $a_h$ in these integrations.

We conclude that the binary hardening rates
observed in the $N$-body simulations -- including
their dependence on $M$ and $N$ -- are consistent with
the predictions of loss cone repopulation theory.
In particular, the integrations with $M=0.04$
exhibit the $\sim N^{-1}$ scaling of hardening rate with
particle number that is characteristic of an ``empty,''
diffusively-repopulated loss cone.

Brownian motion of the binary might also affect its
decay rate, by enhancing the diffusion of stars into
the binary's loss cone.
In Paper I, a simple model was presented for this process.
Equation (85) from that paper gives the diffusion
time (i.e. the time for the binary's loss cone to be repopulated)
due to Brownian motion as
\beq
t_{Brown}\approx {(Ka)(2E)^{7/2}\over 6\pi G^2(M_1+M_2)^2{\cal A}^2}.
\eeq
Here $Ka$ is defined as above, as the distance from the binary's 
center within which the gravitational slingshot is effective; 
$E$ is the orbital energy, assuming a Keplerian potential around 
the black hole (the potential from the other stars was ignored
in this model);
and ${\cal A}^2$ is the mean square acceleration experienced
by the binary due to gravitational perturbations from stars.
Adopting equation (76) from Paper I for ${\cal A}$,
and setting $E=\sigma^2$, the orbital energy of a star near the binary's
gravitational influence radius, we find in $N$-body units
\beq
t_{Brown}\approx 10^2 N_6M_{-2}a_{-2}
\eeq
where $N_6\equiv N/10^6$, $M_{-2}\equiv (M_1+M_2)/0.01$, and $a_{-2}\equiv a/0.01$.

This time scale should be compared with the time scale for stars
to diffuse into the binary's loss cone due to star-star encounters,
or $t_{diff}\approx P(E)/q(E)$ with $P(E)$ the orbital period.
Evaluating $q$ and $P$ at the binary's gravitational influence
radius, $E\approx E_h=\sigma^2$, we find
\beq
t_{diff}\approx \left\{0.59,0.15\right\}q(E_h)^{-1}
\eeq
for $M=\{0.04,0.01\}$.
In the simulations with the smaller binary, 
Figure~\ref{fig:qofe} shows that $q$ is always greater than
$\sim 1$, implying $t_{diff}\lap 0.15$.
This is much shorter than $t_{Brown}$ for any $(N, a)$ considered
here, from which it follows that Brownian motion had almost
no effect on loss-cone repopulation.
In effect, the smaller binary's loss cone was maintained in such a full
state by star-star encounters that the additional influence
of the binary's motion was negligible.
In the case of the more massive binary,
we have 
\beq
{t_{diff}\over t_{Brown}}\approx \left({10^3 qN_6a_{-2}}\right)^{-1}.
\eeq
Figures~\ref{fig:ainv} and~\ref{fig:qofe} suggest that this
ratio might have approached unity at late times in the simulations
with smallest $N$, but generally was much smaller than one,
again suggesting that the influence of Brownian
motion on the binary's evolution was small.

\section{Comparison with Other Work}

Our initial conditions were chosen to be identical with those of
Chatterjee, Hernquist \& Loeb (2003) (CHL), who used a 
rather different $N$-body algorithm to follow the evolution of
massive binaries at the centers of Plummer-model galaxies.
These authors present detailed results for only two integrations,
with $M_1=M_2=0.00125$ and $M_1=M_2=0.02$; the latter values match
our more massive binary, and Figure 2 in CHL shows the evolution
of $1/a$, $e$ and $\rho(0)$ for an integration with $N=0.2\times 10^6$.
The evolution found by CHL for $e$ and $\rho(0)$ appears consistent
with what we find here for the same values of $M$ and $N$.
However our results for the binary hardening rate differ 
from theirs, in two respects.
(1) CHL observed binary hardening rates $s=(d/dt)(1/a)$
that were significantly time-dependent, with $1/a$ 
sometimes increasing as weakly as $\sim t^{0.5}$, rather than
the linear dependence observed here.
For instance, their Fig. 2 shows a gradually falling decay rate 
in the case ($M_1=M_2=0.02$, $N=0.2\times 10^6$), with a value
of $1/a$ at $t=250$ of $\sim 360$, compared with our value of
$\sim 500$.
(2) While CHL also found that hardening rates tended to fall with
increasing $N$, they state that ``$s$ falls systematically until
there are roughly $200,000-400,000$ stars, when it stabilizes
to a particular value, $s_0$.''
Unfortunately, CHL provided no plots of $s(t)$ or $s(N)$ 
against which this result could be verified; 
however they did quote some particular
values for $s_0$, e.g. $s_0 \approx 8$ when $M_1=M_2=0.005$.
This is consistent with the value of $s$ that we measure
for large $N$ in our integrations with the same binary mass
({Fig.~\ref{fig:ndepend}), 
but we observe no tendency for the
hardening rate to ``stabilize'' at large $N$, either for this
binary or for the more massive one that we integrated.

CHL compared the results of their integrations with the predictions
of {\it local} theory, via an equation similar to our equation 
(\ref{eq:simple}).
Since that equation predicts no dependence of hardening
rates on $M$ or $N$, CHL invoked the mass dependence
of the binary's Brownian motion to explain their results.
CHL suggested (without presenting a quantitative theory) that
the larger wandering radius in integrations with smaller $M$ and
$N$ would translate into larger rates of interaction between the
binary and the stars.
The ``stabilization'' of the hardening rate which they observed
at large $N$ (and which we do not observe) was attributed to a
kind of Brownian-motion-mediated feedback process, 
in which the binary maintains a constant
supply  rate by modulating  the local density of stars.
However no supporting evidence for this model was presented;
for instance, it was not demonstrated that the central
density was actually regulated by the binary in their $N$-body integrations, 
or that the amplitude
of the Brownian wandering increased with $N$ in the 
manner postulated.

Decay of binary black holes in Plummer-model galaxies
was also investigated by \cite{Hemsendorf:02} (HSS).
Those authors placed two equal-mass particles, $M_1=M_2=0.01$,
at the center of the galaxy and carried out  integrations
with particle numbers 
$N=(0.033,0.065, 0.131)\times 10^6$.
The black hole particles were placed on initially eccentric
orbits at radii $\pm 0.5$, farther from the center than
in our or in CHL's simulations.
Furthermore the models were integrated only until $t=60$, and in the
case of the largest $N$, only until $t\approx 40$.
Comparison with our Figure~\ref{fig:slope} suggests that the
hardening rate of the binary at $t\approx 60$ might be rather 
different from its late-time value.
HSS found a mean hardening rate for all of their integrations
of $s\approx 6-7$ (their Fig. 3), quite consistent with
the early hardening rates found here (Fig.~\ref{fig:slope}), 
especially if one takes into account that their binary mass 
lies midway between our two.
HSS observed only a slight dependence of decay rate on
$N$, but this too is consistent with what we find at early
times (Fig.~\ref{fig:slope}).
HSS observed a much stronger evolution of the binary's
eccentricity, but this may be due to their large
{\it initial} eccentricity, $e\approx 0.8-0.9$.
Eccentric binaries are expected to evolve in the direction
of increasing eccentricity \citep{Mikkola:92,Quinlan:96}.

A number of other $N$-body studies of binary evolution have
been carried out using galaxy models with large cores
\citep{Ebisuzaki:91,Makino:93,Colpi:94}, though
most of these used too few particles to show a dependence of
hardening rate on $N$.
One exception was the recent study by Funato \& Makino (2003),
who considered binaries in \cite{King:66}-model
galaxies with particle numbers as high as $1.0\times 10^6$.
These authors also failed to reproduce the stabilization
of the binary hardening rate at large $N$ found
by CHL.

\section{Implications for Binary Evolution in Real Galaxies}

The strong $N$-dependence of the binary hardening rate
in our simulations makes it dangerous to extrapolate
our  results to real galaxies.
Such an extrapolation may nevertheless be justified
in the case of our more massive binary,
which appeared (Fig.~\ref{fig:qofe}) to be in the ``empty loss cone''
regime.
Massive binaries in real galaxies are also expected to
be in this regime (Paper I), at least in the absence
of any additional mechanisms for loss-cone repopulation,
and the $N$-dependence
of the hardening rate observed here for the more massive
binary might therefore be valid for much larger $N$.

We begin by noting that the degree of evolution, as measured
by the ratio $a_h/a(t_f)$ between initial and final binary
separations, was equal to 
$\sim (21,14,8.8,6.2)$ for $N=(0.05,0.1,0.2,0.4)\times 10^6$
in the simulations with the more massive binary.
If the binary's actual mass was $10^8M_\odot$, the $M-\sigma$ relation
\citep{FF:05} implies a galaxy velocity dispersion of 
$\sigma\approx 180$ km s$^{-1}$, so $a_h\approx 1.7$ pc
and the final scaled separations become 
$(0.025,0.12,0.19,0.27)$ pc.
Coalescence of an equal-mass binary in a Hubble time due to gravitational wave
emission requires $a_h/a\gap 20$ \citep{Living:04},
so the integration with smallest $N$ can be said to have
just reached the gravitational radiation regime.

But the implied degree of hardening in a real galaxy is much less
than this if we take into account the likely scaling of
the hardening rate with $N$.
Ignoring possible changes in the central density
of the host galaxy (such changes were small in the integrations 
reported here),
equation (9a) implies
\begin{mathletters}
\begin{eqnarray}
T_{decay} &\equiv& \left[a{d\over dt}\left({1\over a}\right)\right]^{-1} \\
&\approx& 2.0\times 10^4 \left[G\rho(0)\right]^{-1/2} \left({r_0\over a}\right) 
N^{0.81}
\end{eqnarray}
\end{mathletters}
or
\beq
{T_{decay}\over T_{cr}} \approx 2\times 10^7 \left({r_0/a\over 10^3}\right) 
\left({N\over 10^{11}}\right)^{0.81}
\label{eq:tdecay2}
\eeq
with $T_{cr}\equiv (G\rho(0))^{-1/2}$ the galaxy crossing time.
Equation (\ref{eq:tdecay2}) suggests that binary hardening times in
real galaxies would be very long, of order $10^7$ crossing times
and much greater than the age of the universe.
In effect, the binary would ``stall.''

A very different conclusion was reached by \cite{CHL:03},
who {\it assumed} that the binary hardening rates observed
by them at $N\approx 0.2-0.4\times 10^6$ would remain
constant even for much larger values of $N$.

Of course our prediction is based on a rather unphysical
(too homogeneous) galaxy model, 
an unrealistic (too large) choice for the mass of the binary, 
and unlikely initial  conditions (with the two black holes
deposited symmetrically about  the center of the galaxy,
rather than arriving there as a result of a galactic merger).
Binary hardening rates in galaxies with steeply-rising central densities
might be significantly larger than implied by equation (\ref{eq:tdecay2}).
Upcoming papers in this series will attempt to correct these deficiencies
with more realistic galaxy models and more physically-motivated
initial conditions.

\section{Conclusions}

The evolution of a massive binary at the center of a galaxy with
a large, constant-density core was followed using a high-accuracy
$N$-body code and a special-purpose computer cluster.
Two values for the binary mass were considered: 
$(M_1+M_2)/M_{gal}=(0.04,0.01)$.
The binary's hardening rate exhibited a clear $N$-dependence,
consistent with the predictions of collisional loss-cone repopulation
theory.
In the simulations with $M_1+M_2=0.04M_{gal}$ and large particle number, 
$N\gap 0.2\times 10^6$, the binary appeared to be nearly in 
the ``empty loss cone'' regime believed to characterize
real galaxies
(in the absence of other mechanisms for loss-cone refilling):
the hardening rate decreased almost as steeply as $N^{-1}$.
We observed no indication that the binary hardening rate ``stabilizes'' at
particle numbers greater than $N\approx 0.2\times 10^6$,
or that the binary's evolution is strongly affected by its Brownian motion,
as claimed in a recent study based on
a more approximate $N$-body treatment (Chatterjee, Hernquist \& Loeb 2003).

Our results demonstrate the feasibility of $N$-body simulations of 
binary black hole evolution that mimic the behavior expected
in systems of much larger $N$.
Even larger particle numbers, $N\gap 10^6$, will be required to
correctly reproduce  the evolution of binary black holes 
in more realistic galaxy
models with higher central densities and less massive binaries.
Direct-summation $N$-body simulations with such high particle
numbers are now becoming feasible with the advent of parallel, 
special-purpose computers like gravitySimulator.
Future papers in this series will present the results of such
simulations.

\acknowledgments

DM acknowledges support from grants
AST-0206031, AST-0420920 and AST-0437519 from the 
NSF, grant NNG04GJ48G from NASA, and
grant HST-AR-09519.01-A from STScI.
PB and RS acknowledge support from
grant SFB-439 from the Deutsche Forschungsgemeinschaft.
This work was supported in part by the Center for Advancing the Study of
Cyberinfrastructure at the Rochester Institute of Technology.

\end{document}